# Crystallisation From Volatile Deep Eutectic Solvents

*Jason Potticary[1], Charlie Hall[1,2], Victoria Hamilton[1,3], James F. McCabe[4], Simon R. Hall[1]\**

[1]Complex Functional Materials Group, School of Chemistry, University of Bristol, Bristol BS8 1TS, UK.

[2]Centre for Doctoral Training in Condensed Matter Physics, HH Wills Physics Laboratory, Tyndall Avenue, Bristol, BS8 1TL, UK.

[3]Bristol Centre for Functional Nanomaterials, Centre for Nanoscience and Quantum Information, Tyndall Avenue, Bristol BS8 1FD, UK.

[4]Early Product Development, Pharmaceutical Sciences, IMED Biotech Unit, AstraZeneca, Silk Road Business Park, Macclesfield, Cheshire SK10 2NA, UK.

A new class of deep eutectic solvents are presented which exhibit all of the physical characteristics of classical deep eutectic solvents, with the exception that one of the components is volatile when exposed to the atmosphere at room temperature. This enables a premeditated, auto-destructive capability which can lead to novel crystalline identities. We demonstrate the effectiveness of this concept through the room-temperature crystallisation of a broad range of organic molecules, with a particular focus on pharmaceuticals, that possess a variety of functional groups and molecular complexity. Furthermore, we show how, through the simple altering of the eutectic composition, polymorphism in paracetamol can be controlled, enabling the elusive metastable form II to spontaneously crystallise at room temperature.

Born from the class of solvents known as ionic liquids, the deep eutectic solvents (DES), named from the Greek "εu" (eu = easy) and "τήξις" (teksis = melting), have been an increasingly well-researched class of solvents for the last two decades. They have been a boon to catalysis, extraction processes, electrochemistry, organic synthesis and in the creation of more efficient batteries and dye-sensitized solar cells.[1-7] Originally conceived as a greener and cheaper alternative to the more toxic and less environmentally friendly ionic liquids,[8,9] DESs are normally composed of two or three molecular species which are able to associate through extensive hydrogen bonding[10] bringing about their inherent stability. The judicious admixture of two solid crystalline components forms a liquid with a melting point significantly lower than each of the constituents giving rise to their ubiquity. One of the first examples of a deep eutectic solvent was formed by the mixture of solid choline chloride and urea in a molar ratio of 1:2. This eutectic mixture is a liquid at 12 °C, whilst the component parts have melting points of 302 °C and 133 °C respectively. Although the first DESs were formed by the mixing of a quaternary ammonium salt with a hydrogen bond donor (HBD),[11] a number of suitable alternative formulations have subsequently been conceived, the majority utilising an amide moiety as a hydrogen bond acceptor (HBA). DESs are being investigated within the pharmaceutical industry to tackle one of its biggest issues, that of poor solubility. Currently, more than 50% of active pharmaceutical ingredients (APIs) are known to exhibit polymorphism and up to 90% of new chemical entities considered for bringing to market are classed as poorly soluble.[12] DESs have been explored both as a vector for enhancing the solubility of an API and the incorporation of an API as one of the eutectic components, forming "therapeutic DESs",[13-16] although the bioavailability of APIs delivered in this form has been shown to be lower than standard oral solids.

      When developing an API, one of the primary considerations for bringing a new chemical entity to market is the complete understanding of its polymorphic landscape. A sudden change in polymorphic form can result in radical changes to intermolecular interactions and the crystal surface chemistry which can detrimentally affect solubility, dissolution rate and intestinal permeability.[17] On a larger scale, different polymorphs often have different crystal habits which can have a significant impact on the processability of the API. Greater cohesion between crystals from the increased amount of exposed polar surfaces, leads to clogging of processing hardware. Typically, additional mechanical processes such as milling are used to reduce particle size. However, milling is accompanied by concerns over mechanically induced solid-state

transformations of the API and agglomeration of the produced particles.[18,19] For these reasons, identification of new polymorphs alongside control over polymorphism, crystal habit and micromorphology remain major resource sinks within the pharmaceutical industry. The socioeconomic and research impact of directing crystal growth in pharmaceutical compounds cannot be overstated; existing drugs which have no scalable route to a soluble form with a desirable crystal habit encounter delays in development, because they are not easily processable into tablets with a suitable dissolution profile. Unfortunately, the most thermodynamically stable polymorph of an API is the least soluble when compared to higher energy metastable forms. However, these higher energy polymorphs often require relative extremes of crystallisation conditions or complex synthetic routes such as desolvation or epitaxial growth to isolate. This is the case with one of the most widely produced APIs, paracetamol,[20] which has three known polymorphs, form I, II and III, named in order of discovery. Paracetamol is manufactured and distributed as form I (Figure S7a), a less efficacious form, due to its ease of production and crystalline stability. Form II (Figure S7b) is more soluble and is more readily compressed into tablets, however it is more challenging to crystallise, requiring the addition of additives[21,22] controlled cooling from a melt[23] or bespoke solvent mixtures.[24] Using existing methods, obtaining higher energy polymorphs of any API is not a cost-effective strategy when having to scale up and therefore what is required is a simple method of API crystallisation and polymorph control which is scalable and works at room temperature and pressure.

Here we introduce a new class of deep eutectic solvent which, through spontaneous reversal of the eutectic liquidation, affords the potential for either (i) polymorphic control, (ii) macromorphological control (iii) or the creation of a novel solvate of an API, depending on the API's molecular structure. This concept of a self-destructing deep eutectic solvent has not been previously explored. We show the efficacy of these volatile deep eutectic solvents (VODES) by crystallisation of APIs via a liquid-based method which is scalable and works at room temperature and pressure. Furthermore, we discuss the molecular features of the API that give rise to either polymorph control, macromorphological control or solvate formation and suggest general features that should make our method applicable to the controlled crystallisation of all small organic molecules.

**Volatile deep eutectic solvent synthesis and polymorphic control in paracetamol**

The creation of a VODES involves a volatile HBD and stable HBA component (in this work the ratios 1:1-10:1 - HBD:HBA, respectively were used) which, when simply mixed together as solids, produces a liquid (**Figure 1(a)**) which remains stable in a sealed container at or near room temperature. This admixture is subsequently left to 'self-destruct' at room temperature and pressure for a time, $T_x$ (typically ~36 hr), resulting in the spontaneous crystallisation of the non-volatile component. In this work, we use a pharmaceutical as the HBA component, which means that in lieu of dissolution of and concentrations in a solvent, the liquid produced, is itself, part API which in some cases comprises 20% of the liquid. In addition, given that we are applying our method to pharmaceuticals, it is of the utmost importance that the amount of phenol present in the crystalline product is within FDA allowed limits for use in humans. In this work we therefore undertook solution $^1$H-NMR of each as-crystallised product and in every case, there was < 0.25% w/v residual phenol present (Figures S2 - S6). This is below the FDA allowed limit for oral ingestion, topical application and by injection (intradermal, intralesional, intramuscular, intrasynovial, intravenous, or subcutaneous).[25]

In order to demonstrate the viability of the VODES method, we use a simple system consisting of phenol as the HBD and paracetamol as the HBA. Phenol and paracetamol were chosen as simple molecules containing functional groups which are typical in DES formation. In particular, phenol and phenolic compounds have been widely explored as the hydrogen bond donating component in eutectic systems for both extraction processes and pharmaceutical applications.[13] With this system, we are able to actively control polymorphism in paracetamol by varying the HBD:HBA ratio. Regioisomers of paracetamol, where the acetamide group can be on three possible ring positions, para-, meta- and ortho- (PAP, MAP, OAP respectively) (Figure S1) were investigated. Of the two other regioisomers, MAP is considered a potential less-toxic alternative to PAP and has two known polymorphs. One, a catemeric orthorhombic from and the second, a dimer-based, monoclinic form reported recently.[26] OAP is the least studied of the '-cetamol' family, with only a single crystal structure recently solved via electron diffraction.[27] OAP has no current industrial or pharmaceutical applications but has shown potential as an anti-tuberculosis therapeutic[28] and as a highly promising candidate in the treatment of arthritis.[29] Using the VODES method, PAP, MAP and OAP were made into homogeneous liquid VODESs using phenol as the HBD. PAP and MAP were found to produce a stable VODES between the ratios of 4:1 - 9:1 and 5:1 - 10:1 (HBD:HBA), respectively. OAP formed no stable deep eutectic up to a ratio of 10:1. This shows clearly that the position of the acetamide relative to the hydroxyl group is playing a pivotal role in how the VODES forms. When in the para position there are a wider range of approach angles available for the phenol to hydrogen-bond with the API, thereby enabling the formation of a stable VODES at lower HBD:HBA ratios. In contrast to this, when in the ortho position in OAP, the proximity of the hydroxyl group to the amine precludes hydrogen bonding around the molecule in numbers sufficient to make a stable liquid. VODES systems created from PAP and MAP however are so easily formed that they allow crystals to be grown from a solution with API concentrations inaccessible by many common organic solvents. We demonstrate that in the case of PAP, the most efficacious polymorph (form II) currently considered commercially unsuitable for production due to difficulties discussed above, will emerge from a VODES spontaneously at room temperature and pressure. Evaporation of the volatile phenol component from all HBD:HBA ratios, leads to the formation of PAP polymorphs over a 2-day period. Analysis of resultant crystals from phenol:PAP ratios from 4:1 - 9:1 show a stark difference in crystal habit, usually between the ratios 6:1 and 7:1. Powder X-ray diffraction analysis shows that this difference is due to different polymorphs being formed, namely form I (4:1 - 6:1) and form II (7:1 - 9:1) of which each crystal habit observed is characteristic; diamonds in the case of form I and feather-like needles in form II (Figure 2). The selection of form I or form II paracetamol as a simple function of a change in VODES composition at room temperature is a completely reproducible and non-stochastic method of polymorph control. We discuss potential mechanisms underlying this control in the section below.

In contrast to PAP, MAP retains a single crystal structure throughout all VODES composition ratios tested however there is still a stark division between two crystalline morphologies upon VODES destruction. There is a statistically clear divide between the ratios of 7:1 and 8:1 (phenol:MAP) at which the morphology changes from rods to silky needles. Regardless of morphology however, in all experiments, only a single polymorph of MAP, form I, is observed (Figure 3). The final regioisomer, OAP, formed no deep eutectic solution and so no crystals were generated.

What is deeply intriguing is that there is a VODES ratio in PAP where one polymorph consistently forms and not the other, despite the fact that as the phenol evaporates, a VODES of higher concentration must presumably pass through the same phenol:API ratio as one of a lower ratio that gives a different polymorph. For example, phenol:paracetamol of a ratio of 8:1 consistently produces form II, despite passing through an (e.g.) 5:1 ratio as the phenol evaporates, before crystallising. Starting with a phenol:paracetamol ratio of 5:1 however consistently produces form I. This phenomenon can be understood to be indicative of the fact that for each starting ratio, the HBD:HBA interaction that is established from the start is likely to be subtly different from each other. It follows therefore, that from the start of the evaporation of the phenol, both systems contain distinct molecular environments and would therefore not be expected to undergo crystallisation in the same way. Each system may possess characteristic pre-nucleation clusters from the start, which persist until crystallisation occurs; indeed, such clusters have recently been observed in APIs.[30]

**Understanding the eutectic behavior of benzamides**

With the knowledge that the number and propensity of HBD:HBA interactions are critical to the successful formation of a VODES, the benzamides are appropriate pharmaceuticals to serve as model systems, as this structural motif is common to many drugs.[31] Benzamide itself currently has three known forms (I, II and III) with the highly metastable form II and the less metastable form III has, to date, only been reported to crystallise concomitantly at higher supersaturations from standard solvents, either with one another, alongside the stable form I or as a trimorphic system. Form III has, however, recently been obtained through solvent-assisted grinding of form I benzamide and form I nicotinamide.[32] The effect of altering functional groups on the eutectic behavior was examined using 2-methoxybenzamide (2MB) and 2-ethoxybenzamide (2EB), as simple variations on the underlying benzamide structure. Both 2MB and 2EB have only a single known polymorph, but it is of note that while 2MB possesses a catameric bonding motif, 2EB has a dimeric motif between the amide substituents. A dimer-motif is also present in the two room-temperature stable polymorphs of benzamide, forms I and III.

Phenol:benzamide mixtures within the range 3:1 - 10:1 all resulted in a homogeneous, clear liquid being produced at room temperature. Thermal analysis determined the lowest recorded melting points of benzamide, 2MB and 2EB phenol-VODESs to be -32.89 °C (5:1), < -70.00 °C (4:1) and < -70.00 °C (4:1), respectively (Table S1). On the destruction of the benzamide VODES, crystals either large, clear, plates of form III or a mixture of form III interspersed with form I benzamide were formed. The polycrystalline nature of the form I benzamide crystals implies that the sporadic formation is due to the formation of a metastable phase (Figure S8) which then rapidly transforms to form I benzamide. 2MB exhibits no unusual behaviours, easily forming a VODES within the standard range of ratios and subsequently undergoing a complete separation to recrystallise as its sole known polymorph. Similarly, 2EB generally undergoes a crystallisation to the known crystalline form, however, the formation of a stable 2EB-phenol co-crystal is observed in a small number of crystallisations. This novel co-crystal comprises a 1:2 (Phenol:2EB) stoichiometry of the two co-formers with the 2EB molecules arranged in a carboxamide dimer with the phenol interacting peripherally (Figure S17) (CCDC deposit number 1879336).

An understanding of the root cause in the variation in eutectic behaviours can be gleaned from thermal analysis of the three related benzamides. Thermograms of higher ratios of VODES containing benzamide show complex features with multiple thermal events on warming that suggest the benzamide-phenol mixture has at least one hitherto unisolated metastable phase (Fig S11). This complex behaviour is reflected in the thermograms of 2EB, in which case the metastable phase is known to be the 2EB-phenol co-crystal (Figure S13), allowing it to be inferred that the transient phase present in benzamide is a benzamide-phenol co-crystal phase of unknown stoichiometry. The thermograms of benzamide and 2EB are in stark contrast to the comparatively featureless thermograms of 2MB which have only a single crystallisation phase change (Figure S12). This suggests that the crystallisation behaviour observed from the benzamide and 2EB eutectic systems is related to the ease of incorporation of phenol molecules within their dimeric structural motifs, which differ from the catemeric motifs present in 2MB. It is likely that the formation of the difficult to obtain pure form III benzamide results from cases where the transient phenol structural motifs lead to the inclusion of phenol molecules into the lattice, as has been reported for seeded growth utilising nicotinamide molecules,[32] but in this work, the 'seeds' (in the form of phenol molecules) are able to escape the system due to their volatility. These results indicate that in the absence of an interim phase, traditional supersaturated crystallisation will take place as the phenol leaves the system, which is what we observe for 2MB.

**Thermal analysis and applicability**

To understand the extent of melting point depression in these deep eutectic solvents and to determine the presence of intermediate phases, DSC was performed on all compositions and ratios of phenol:API between 1:1 – 10:1 that produced a stable liquid (Figs. S9 – S16). These data appear system specific with regards to the extent of melting point depression and overall physical behavior of the mixture. Although distinct endotherms, exotherms or glass-transitions were not always present, all VODES systems show large melting point depression for all molar ratios studied (Table S1). When present, melting endotherms are typically broad and often occur at or just after the crystallisation exotherm, particularly at lower phenol:API ratios, leading to a ± 5 °C uncertainty in melting point depression.

In the case of PAP there is no indication of a difference in thermal behaviour as the phenol:API ratio is changed, highlighting that the intermediate phases that lead to the resultant polymorph are not necessarily crystalline in nature and are forming in the liquid state. To evaluate how the differences in pre-nucleation dynamics may change with phenol:API ratio, molecular optimisation calculations were carried out in ORIENT,[33] which suggests how the change to the more planar form II paracetamol allows for more phenol molecules to solvate the crystal as it grows (Figure 4). Conversely however, in MAP, where a difference in morphology could be seen between ratios of 7:1 and 8:1 phenol:MAP, an additional crystallisation and melting event can be seen occurring for solutions above 7:1. This may be an indication that the change in morphology is again due to a short-lived intermediate metastable phase and certainly suggests that the bonding motifs within the solution are a function of the phenol:API ratio.

The successful formation of a VODES was observed in a range of pharmaceuticals. Table S1 describes a list of all APIs used as HBAs during this study. Upon attempting VODES creation with some of these molecules, it is clear that it is not simply a question of propensity to hydrogen-bond which determines the eventual crystalline form. In some cases, formation of a stable novel co-crystal of API and phenol occurs, with the phenol effectively playing the role of 'solvate' in the crystal. In the case of harmine, a reversible monoamine oxidase inhibitor, the phenolate is a precursor to the only know native crystalline form. The harmine phenolate has a stoichiometry of 1:1 (HBD:HBA), with the hydroxyl group on the phenol hydrogen-bonding with the amine on the harmine molecule and has been solved by us (Figure S18 - CCDC deposit number 1879689). Upon gentle heating or vacuum drying of the harmine phenolate, the macromorphology of the crystal is preserved (Figure 5(a, b)), but the large single crystals have been transformed into a porous, polycrystalline matrix through loss of phenol (Figure 5(c)). The production of a porous API through the evaporation of solvent has been demonstrated previously for clozapine to which the porous structure increased the dissolution rate 5-fold.[34] Both carbamazepine and verapamil hydrochloride show no crystallisation or melting events at any ratio except an exotherm followed directly by and endotherm in the carbamazepine system at 10:1. Both systems display a $T_g$ on both cooling and warming, which moves systematically towards lower temperatures with higher ratios. Metaxalone (Figure S16) only has distinct crystallisation and melting events with higher ratios of phenol, implying the formation of a co-crystal with a large excess of the HBD. Glass transitions are seen in all ratios tested, including those with crystallisation/melting events.

Overall, thermal analyses of each VODES reveals varying levels of complexity from system to system. Where carbamazepine and verapamil hydrochloride appear as simple glasses, other systems like benzamide show multiple crystallisation and melting events, implying the formation of unknown crystalline forms of either each component or co-crystals of varying stoichiometries and transitions from form to form. This is suggestive that the effects seen, polymorphlogical or morphological, are a function of the strong interactions with volatile component of the VODES. Systems that have had transient co-crystals isolated and solved appear to demonstrate this and others that have had metastable crystals observed during crystallisation do indeed appear to transform in situ leaving the ultimate polymorph.

Interactions between components in a VODES are dynamic and are facilitated by an array of possible bonding motifs. For example, the inception of crystallisation of a 1:1 co-crystal dramatically changes the concentrations of the liquid in the system, which must lead to a change in the nature and number of molecular interactions throughout the whole system. This complexity will be manifest in systems even where no co-crystal has been detected at ratios beyond 1:1. The mere fact that some co-crystals have been isolated and solved, provides a likely route for the structural control exhibited by some of these systems, both polymorphological and morphological when grown out of a VODES.

**Outlook**

In this study, we prepared a range of volatile deep eutectic solvents as binary mixtures of an API and a hydrogen bond donor. The deep eutectic solvent is then left to self-destruct through diffusion of the more volatile component out of the system leaving the less volatile pharmaceutical to crystallise. As we have demonstrated, suppression of the melting point in the optimum deep eutectic ratio leads to a crystallisation temperature typically 40 ˚C below room temperature (60 ˚C below that of phenol) and often greater than 100 ˚C below room temperature. It is clear that the mechanisms of crystallisation from a deep eutectic solution are inexorably linked to both the molecular structure of the API and the packing within the resultant crystals. The propensity for potential hydrogen-bonding in each system appears to be pivotal in determining how likely an API is to form a VODES and at what ratios, before there is crystallisation of either the API at one extreme or the co-former at the other. This is exemplified in the case of PAP, MAP and OAP; as the available locations for hydrogen-bond interactions decreases, so do the concentrations of API in the solvent. In the benzamide based systems, spontaneous formation of a stable liquid is again facile with 2-methoxybenzamide, where there are both a carboxamide and ester moieties present, in contrast to benzamide and 2-ethoxybenzamide, whose ester is more sterically hindered. The resultant crystals show that there is a rich and complex interaction between the API and HBD which is system specific.

We have shown here that volatile deep eutectic solvents afford a novel route of crystallisation from liquids of APIs at high concentrations at room temperature. This new crystallisation route provides a potential vector for polymorph selectivity in APIs simply through a change in concentration. When this route is achieved via accessing intermediate solvates, it may also lead to the creation of new porous macromorphologies, thereby enabling the kinetics of dissolution to be altered. These volatile deep eutectic solvents have opened the door to the possibility of controlling organic crystal growth in a scalable way at room temperature, with the additional possibility of creating previously unforeseen polymorphs and solvates. The control over polymorphism and/or the creation of novel polymorphs and solvates in this way has immediate and far-reaching implications for the pharmaceutical industry, where the successful granting and defending of patents relies on having full knowledge of the crystallisation landscape of any given API.

**Experimental Section**

All VODESs in this work were made using ratios - HBD:HBA, which when mixed together as solids, spontaneously produce a liquid (Figure 1(a)) up to 10:1. Notably low ratios (typically 1:1 – 2:1), although becoming paste-like, did not form a homogenous liquid. Once completely liquid, droplets of the VODES were left under ambient conditions for the HBD to evaporate resulting in destruction of the VODES and spontaneous crystallisation of the pharmaceutical HBA. The range of ratios at which a stable VODES is formed affords an easily tunable array of concentrations with regards to the API in the solvent and it is a feature of the VODES system that all eutectic mixtures formed exhibited deep eutectic behavior with melting point depressions or glass transitions significantly lower than the component parts ranging from ~ 29 ˚C to sub -70 ˚C (Table S1).

Methanol and ethanol were supplied by Fisher Scientific (UK). All other chemicals were purchased from Sigma Aldrich and used as received.

**Characterisation**

DSC analyses was performed using a TA Instruments Q2000 and Discovery 25 both with refrigerated cooling systems. The DSC cell was purged with nitrogen. Samples were analyzed in hermetically sealed aluminum pans. A small volume of liquid (~ 2-5 uL) was added to the hermetic pan and sealed. All liquid samples were cooled from 25 °C to -70 °C and then cycled to 70 °C all at 10 °C/min. The instruments were calibrated using a pure indium standard.

Novel crystal structures were solved using single crystal X-ray diffraction. Crystals that were of sufficient size (> 0.2 x 0.2 x 0.2 mm) were isolated during VODES destruction and mounted on a glass fibre. Intensity data for crystal structures were collected at 100 K on a Bruker Kappa Apex II diffractometer, using molybdenum Kα radiation ($\lambda$ = 0.7136 Å). Data were processed using Bruker APEX2 v2.0 and Olex2 v1.2.7 software. For all structures a symmetry-related (multi-scan) absorption correction was applied. Structure solution followed by full-matrix least squares refinement was performed using the WINGXv2014.1 suite of programs throughout.

Specific information about data collection and reflections can be found embedded in the associated cif files.

Powder X-Ray diffraction data were acquired using a Bruker D8 Advance diffractometer with a PSD LynxEye Detector (Cu-Kα radiation wavelength of 1.5418 Å). Step size was 0.0171°/2θ and step hold time was 1.5 s.

Samples for NMR were prepared by dissolving 50 mg of sample in 0.7 cm$^3$ of deuterated solvent with a tetramethylsilane reference standard and filtered. All NMR measurements were carried out on a JEOL ECS-400.

[CCDC 1879336 and CCDC 1879689 contain the supplementary crystallographic data for this paper. These data can be obtained free of charge from The Cambridge Crystallographic Data Centre via www.ccdc.cam.ac.uk/data_request/cif.]

**Acknowledgements**

S.R.H., J.P., C.H. and V.H. acknowledge the Engineering and Physical Sciences Research Council UK (grants EP/L016648/1 and EP/L015544/1), MagnaPharm, a collaborative research project funded by the European Union's Horizon 2020 Research and Innovation programme (grant No. 736899) and the Bristol Centre for Functional Nanomaterials and the Centre for Doctoral Training in Condensed Matter Physics for project funding. S.R.H., J.P., C.H. and V.H. would also like to acknowledge Drs. Hazel Sparkes and Natalie Pridmore for their assistance with single crystal XRD and Dr. Asma Buanz for useful discussions on DSC.

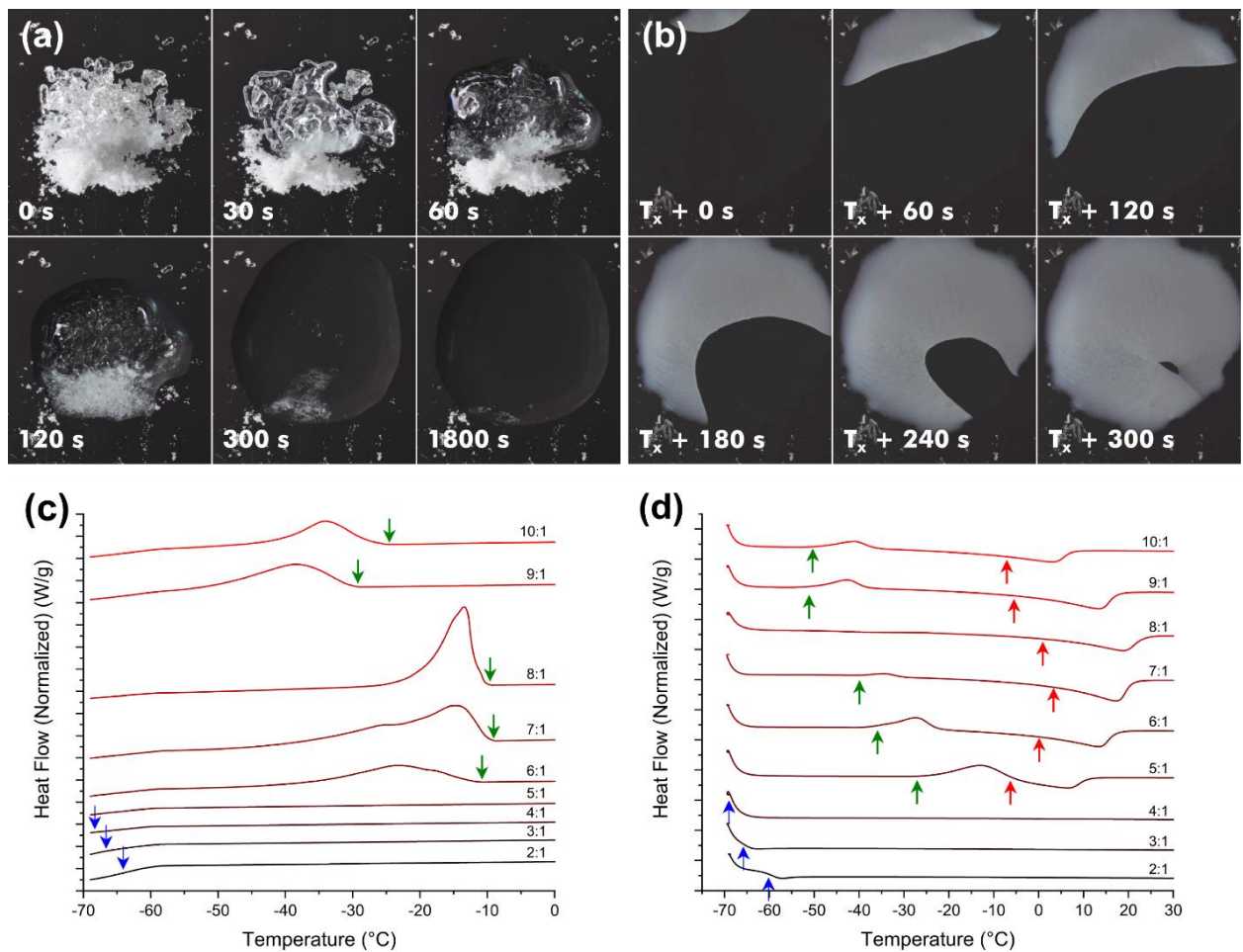

**Figure 1.** Properties of a 2MB-phenol VODES (a) Spontaeneous formation of the VODES at room temperature and pressure. (b) Time-lapse of VODES after onset of crystallisation, ($T_x$ ~24 hr). Thermal analysis of (c) cooling and (d) warming, truncated to remove featureless data. Coloured arrows of blue green and red indicate the positions of a $T_g$, a crystallisation event and a melt event, respectively.

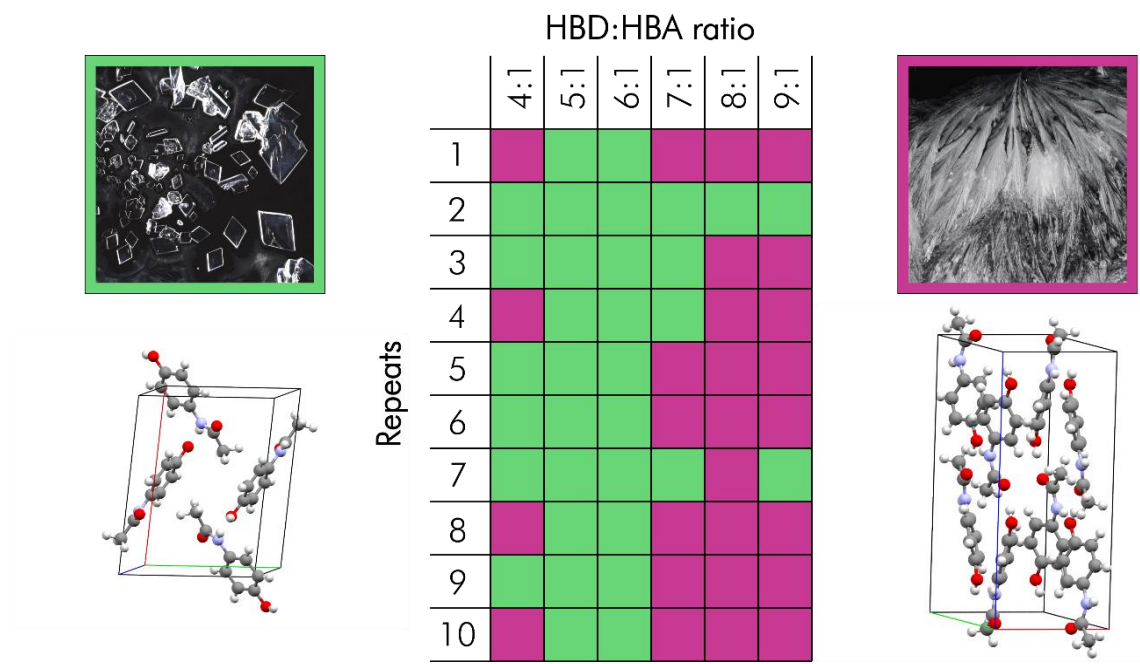

**Figure 2.** Crystalline identity of PAP as a function HBD:HBA ratio. Green and purple squares indicate forms I and II of paracetamol, respectively. Morphology and crystal structures are displayed of form I (left) and form II (right).

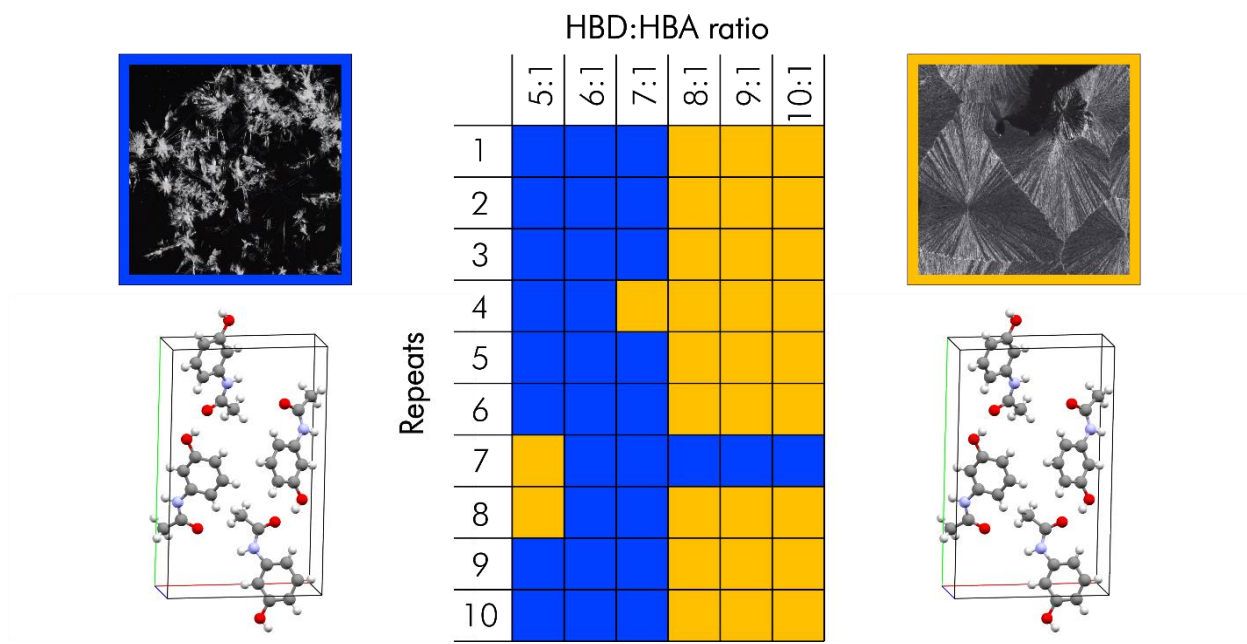

**Figure 3.** Crystalline identity of MAP as a function HBD:HBA ratio. Green and purple squares indicate forms I and II of paracetamol, respectively. Morphology and crystal structures are displayed of form I (left) and form II (right).

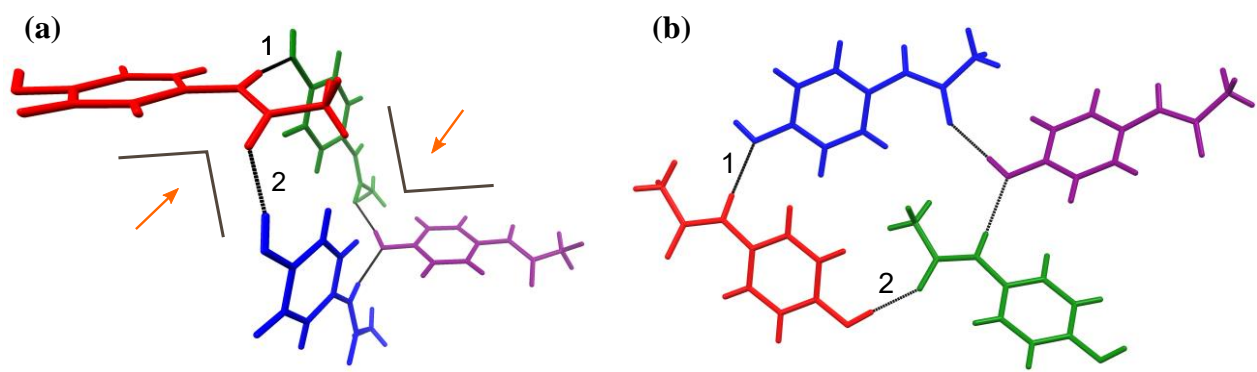

**Figure 4.** Hydrogen-bonding motifs in crystalline forms of PAP. (a) A chair-like cycle of interacting PAP molecules in form I. Orange arrows and taupe cornered lines indicate the angles and sterically obstructed approach vectors. (b) The cycle of interacting PAP molecules in form II showing a more 'planar' structure. Individual molecules are coloured for ease of visualisation. Black lines indicate hydrogen-bonds of lengths (a) 1 = 2.049 Å and 2 = 1.796 Å and (b) 1 = 2.118 Å and 2 = 1.835 Å.

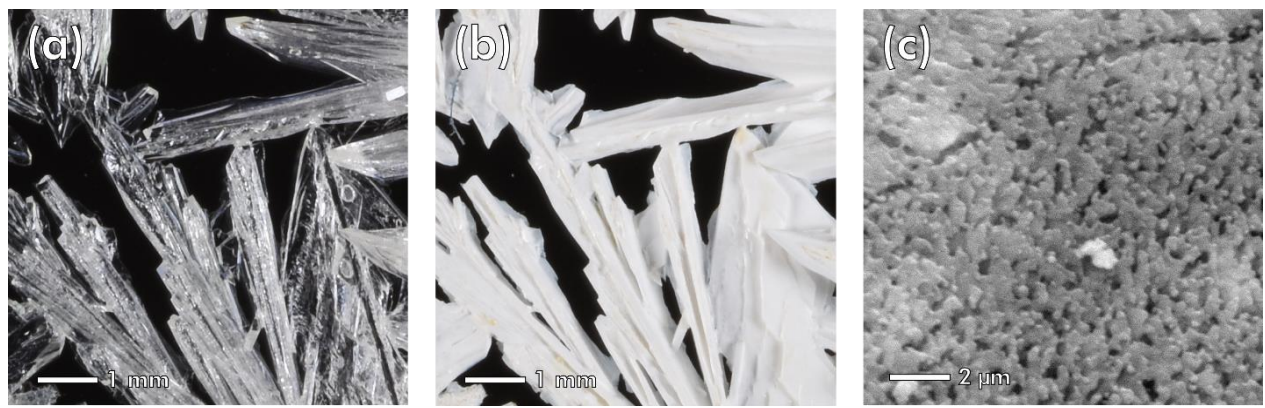

**Figure 5.** Different crystalline forms of harmine. (a) Optical image of crystals of harmine phenolate and (b) an optical image of same crystals after vacuum drying to remove the phenol leaving the standard harmine crystal structure. (c) SEM micrograph of the surface of the dried harmine in (b) revealing a porous micromorphology with an average pore size of 0.22 μm.

**Toc Entry**

Deep eutectic solvents are ubiquitous in the fields of synthetic and materials chemistry owing to their stability, non-volatility and environmentally friendly nature. In this work a new class of them is presented, where one of the two components is volatile when exposed to the atmosphere. Whilst seemingly oxymoronic, the effect of one component leaving the system forces crystallisation with novel results.

**Keyword: Crystallisation**

*Jason Potticary[1], Charlie Hall[1,2], Victoria Hamilton[1,3], James F. McCabe[4], Simon R. Hall[1]\**

**Title**: **Crystallisation From Volatile Deep Eutectic Solvents**

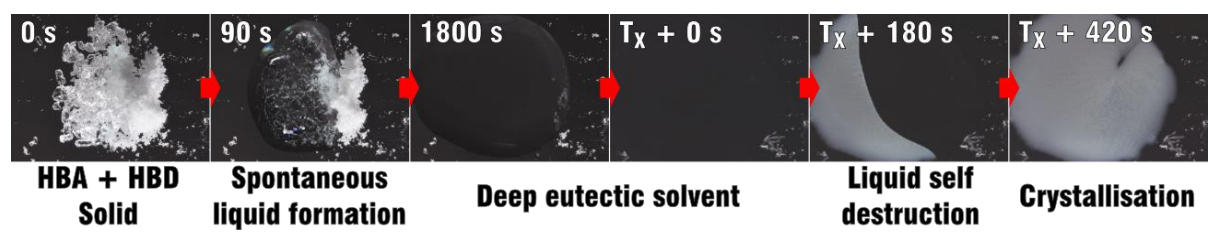

ToC figure

# Supporting Information

## Crystallisation From Volatile Deep Eutectic Solvents

*Jason Potticary[1], Charlie Hall[1,2], Victoria Hamilton[1,3], James F. McCabe[4], Simon R. Hall[1]**

*Correspondence to: simon.hall@bristol.ac.uk

| Compound (m.p. (°C)) | Designation in this study (Figure S1) | HBD (m.p. (°C)) | Lowest VODES m.p. / $T_g$ (°C) | Lowest m.p. ratio (HBD:HBA) |
|---|---|---|---|---|
| Phenol (40.5) | n/a (1) | - | - | - |
| Paracetamol (169) | PAP (2) | Phenol (40.5) | < -70 | 7:1-10:1 |
| Metacetamol (149) | MAP (3) | Phenol (40.5) | -58 | 4:1 |
| Orthocetamol (210) | OAP (4) | n/a | n/a | n/a |
| Benzamide (130) | n/a (5) | Phenol (40.5) | -19 | 10:1 |
| 2-methoxybenzamide (128) | 2MB (6) | Phenol (40.5) | < -70 | 4:1 |
| 2-ethoxybenzamide (134) | 2EB (7) | Phenol (40.5) | < -70 | 4:1 |
| Carbamazepine (192) | n/a (8) | Phenol (40.5) | < -64 | 9:1 |
| Harmine (321) | n/a (9) | Phenol (40.5) | n/a | n/a |
| Metaxalone (122) | n/a (10) | Phenol (40.5) | -58 | 4:1 |
| Verapamil hydrochloride (139) | n/a (11) | Phenol (40.5) | < -70 | 10:1 |

**Table S1.** Summary of melting points of the compounds considered in this study and that of their respective VODES.

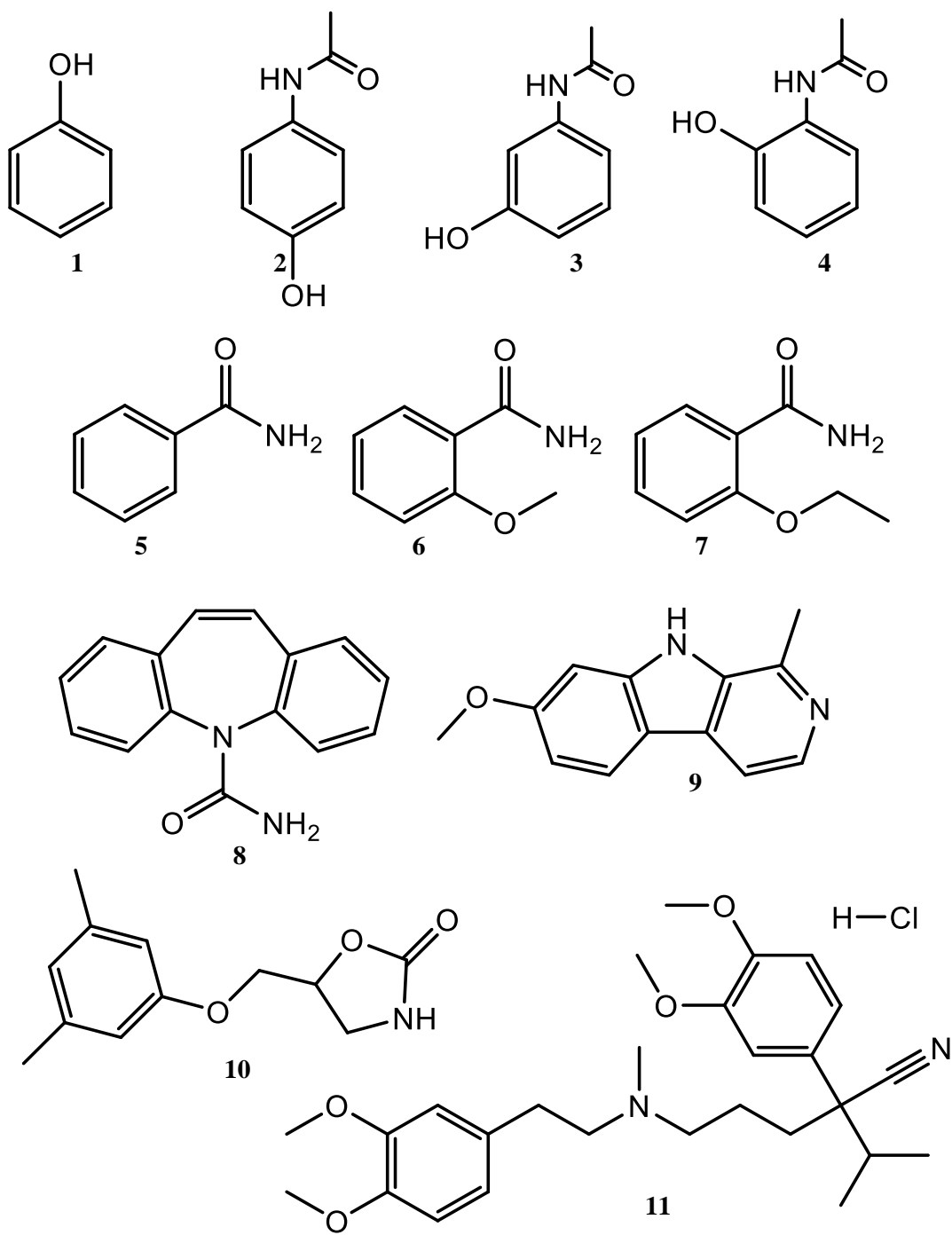

**Figure S1.** The compounds considered in this study. Numbers are identified in table S1.

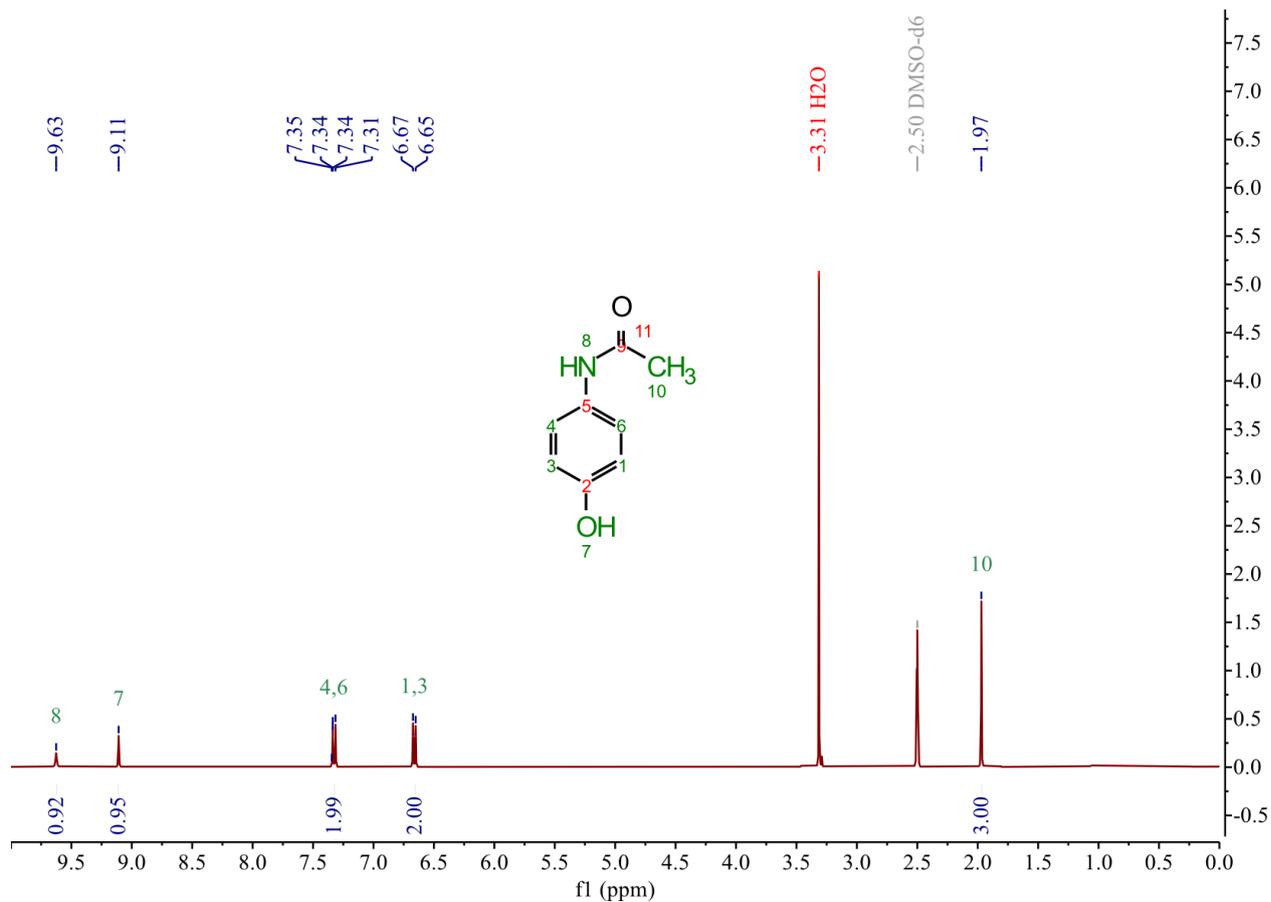

**Figure S2.** $^1$H NMR of paracetamol after crystallisation from a 6:1 VODES with phenol. All peaks can be assigned to paracetamol and are correctly integrated. There is no proton peak from phenol present (c.a. 9.28 ppm in DMSO-d$_6$), indicating an absence of residual phenol in the product.

**Figure S3.** $^1$H NMR of metacetamol after crystallisation from a 6:1 VODES with phenol. All peaks can be assigned to metacetamol and are correctly integrated. There is no proton peak from phenol present (c.a. 9.28 ppm in DMSO-d$_6$), indicating an absence of residual phenol in the product.

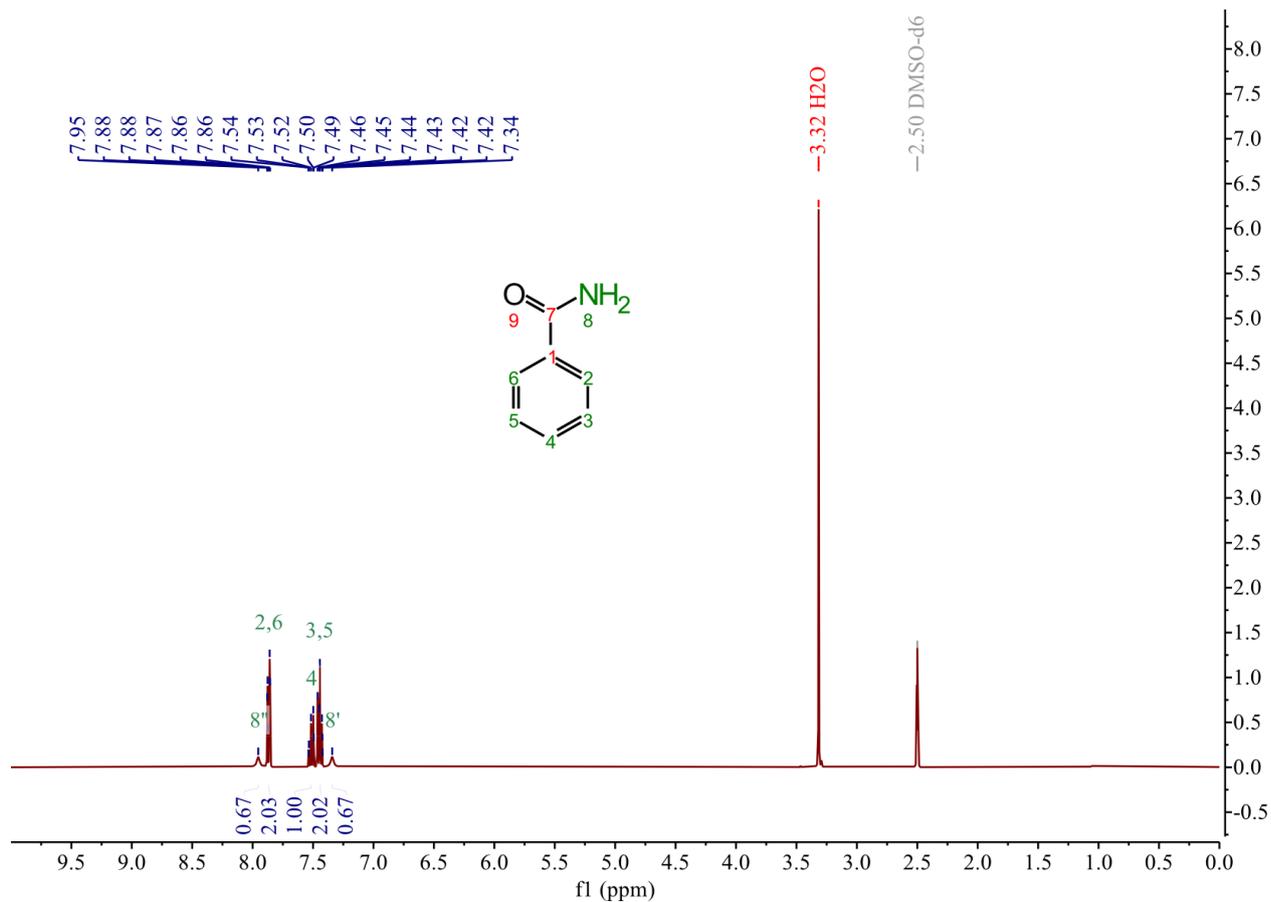

**Figure S4.** $^1$H NMR of benzamide after crystallisation from a 6:1 VODES with phenol. All peaks can be assigned to benzamide and are correctly integrated. There is no proton peak from phenol present (c.a. 9.28 ppm in DMSO-d$_6$), indicating an absence of residual phenol in the product.

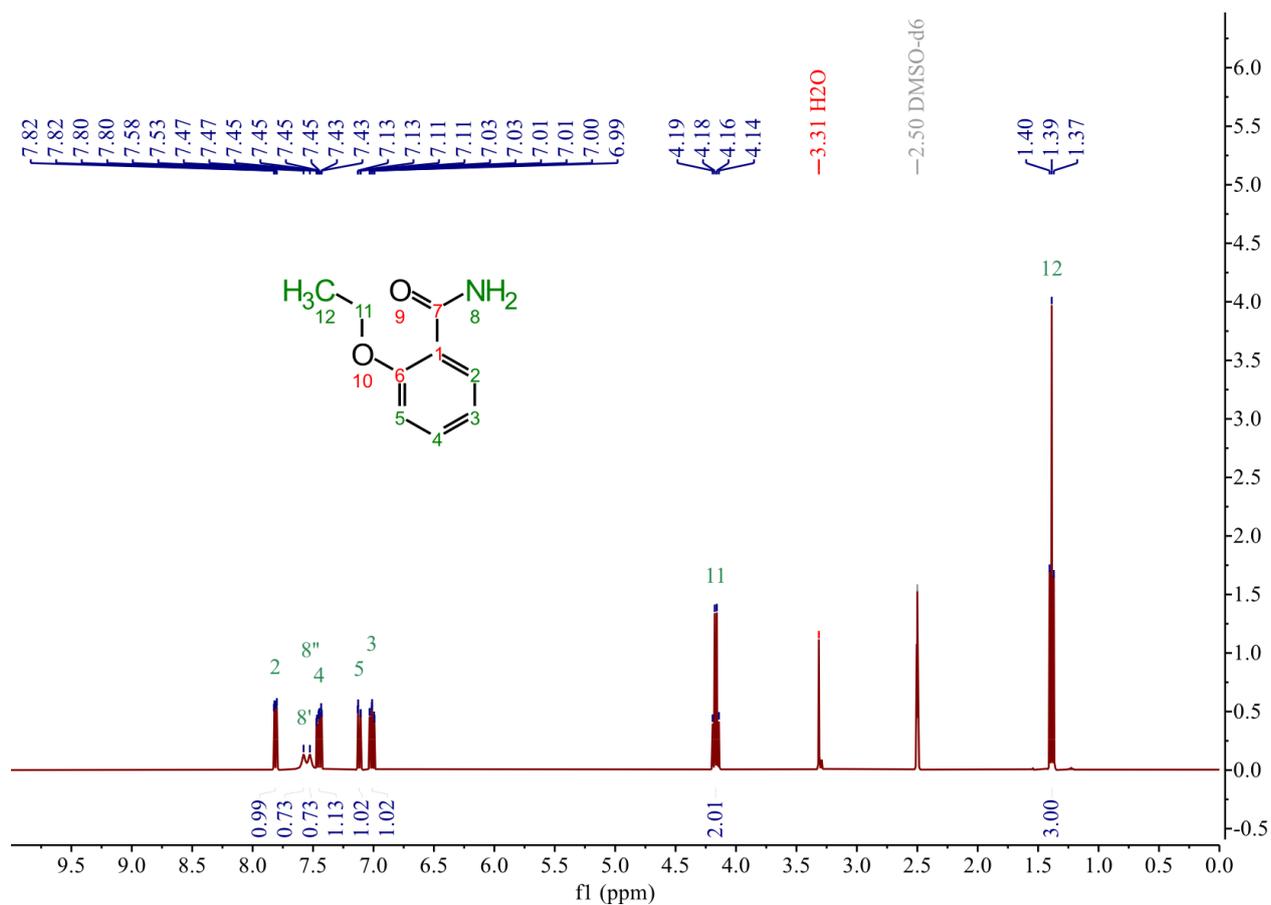

**Figure S5.** $^1$H NMR of 2EB after crystallisation from a 6:1 VODES with phenol. All peaks can be assigned to 2EB and are correctly integrated. There is no proton peak from phenol present (c.a. 9.28 ppm in DMSO-d$_6$), indicating an absence of residual phenol in the product.

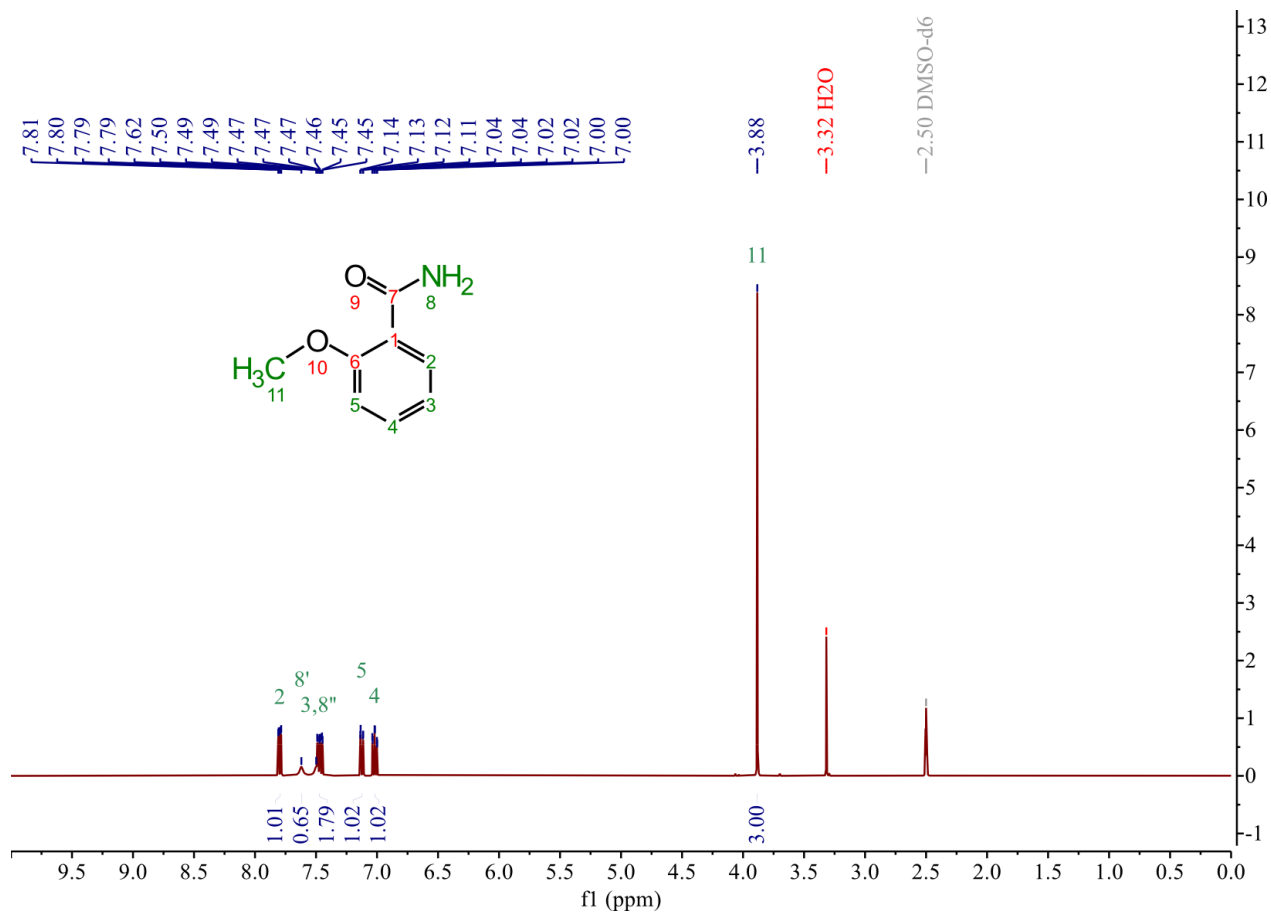

**Figure S6.** ¹H NMR of 2MB after crystallisation from a 6:1 VODES with phenol. All peaks can be assigned to 2MB and are correctly integrated. There is no proton peak from phenol present (c.a. 9.28 ppm in DMSO-d$_6$), indicating an absence of residual phenol in the product.

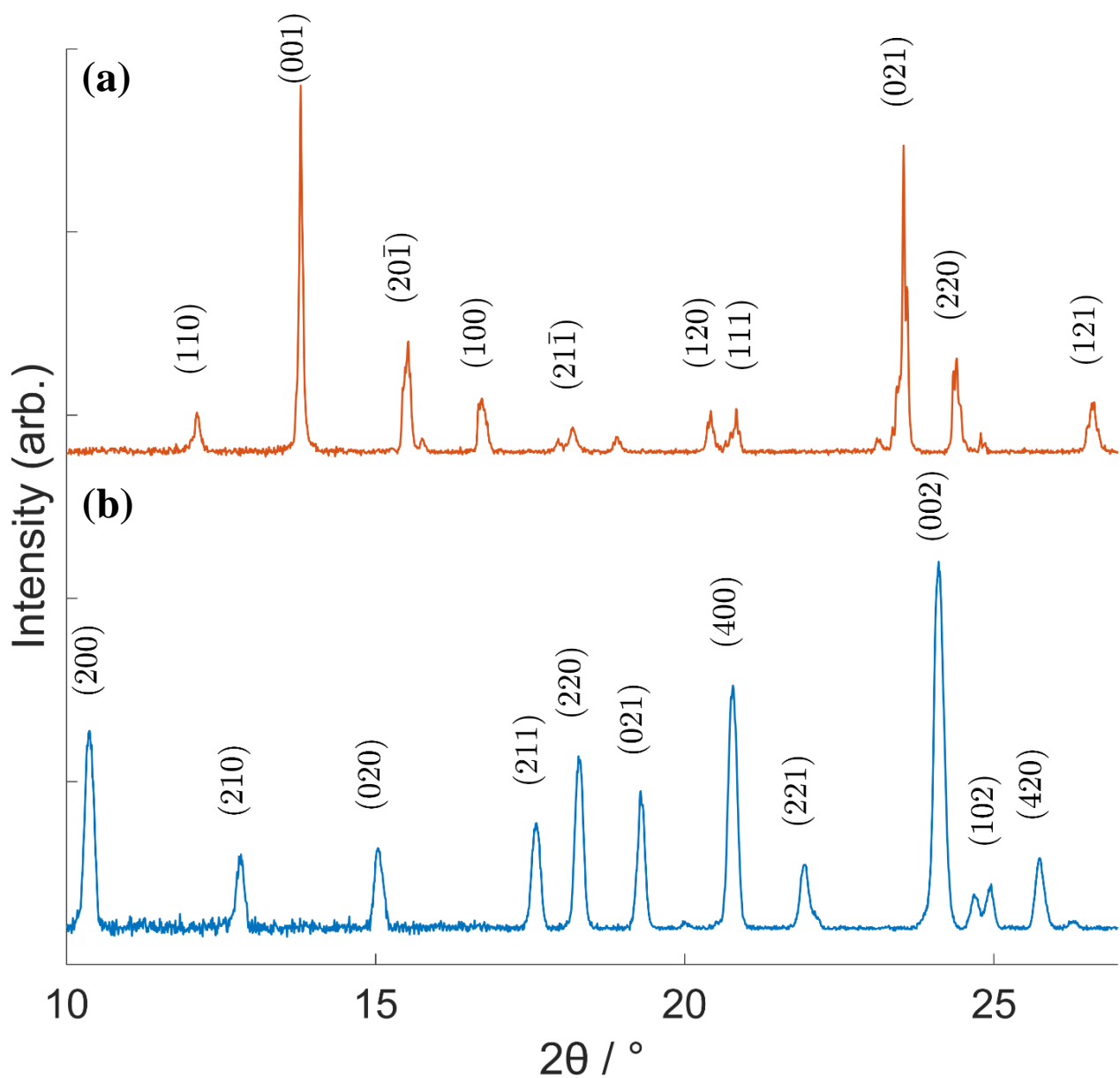

**Figure S7** Powder X-ray diffraction patterns of (a) paracetamol form I (deep eutectic ratios 4:1 - 6:1) and (b) paracetamol form II (deep eutectic ratios 7:1 - 9:1). All peaks can be indexed to their respective polymorph (form I CCDC Identifier – HXACAN01; form II CCDC Identifier - HXACAN).

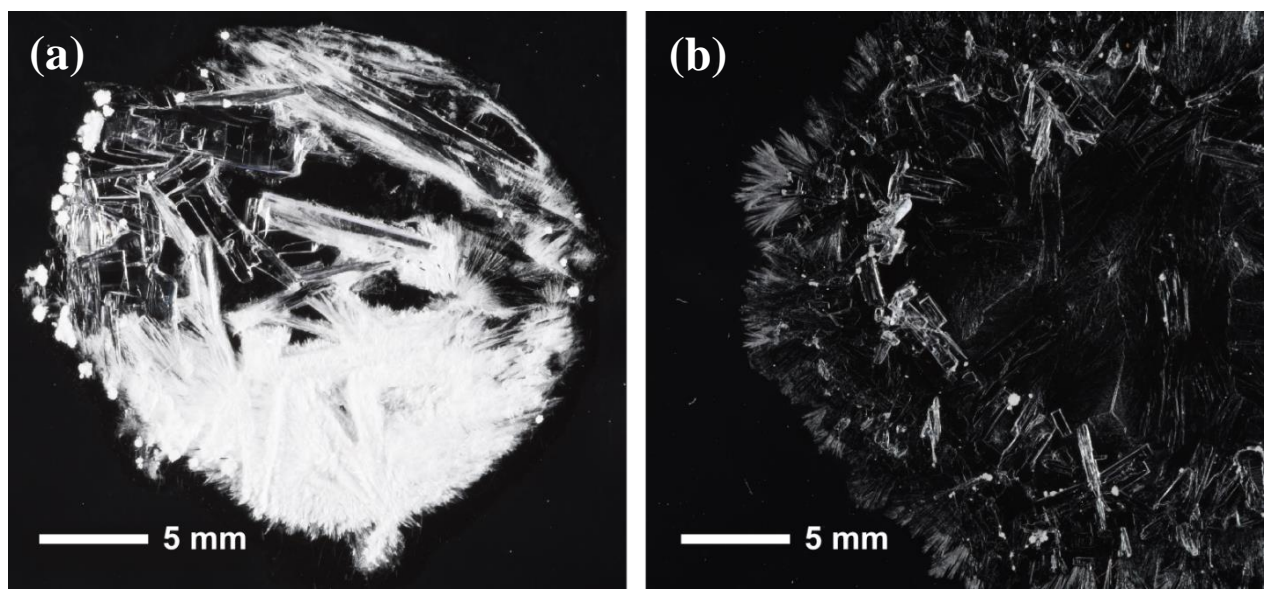

**Figure S8.** VODES crystallisation of benzamide showing (a) both form I (bottom, opaque needles) and form III (top, clear plates) and (b) form III exclusively.

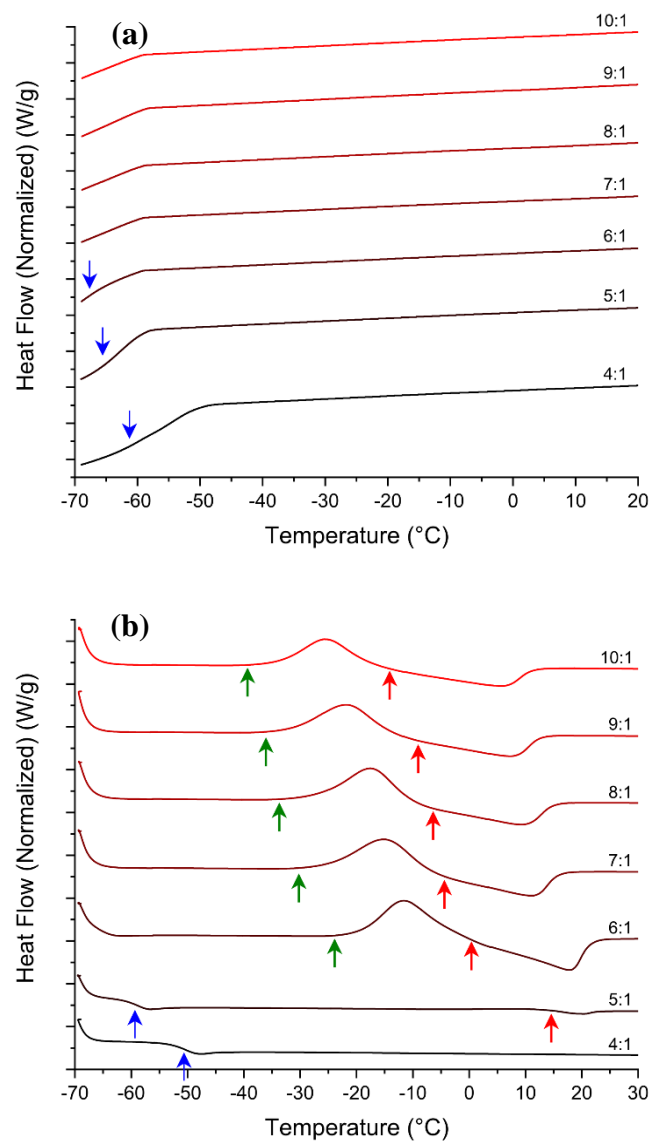

**Figure S9.** Thermal behaviour of paracetamol as a VODES with phenol. Thermograms of each VODES ratio for (a) cooling and (b) warming, truncated to remove featureless data. Coloured arrows of blue green and red indicate the positions of a $T_g$, a crystallisation event and a melt event, respectively.

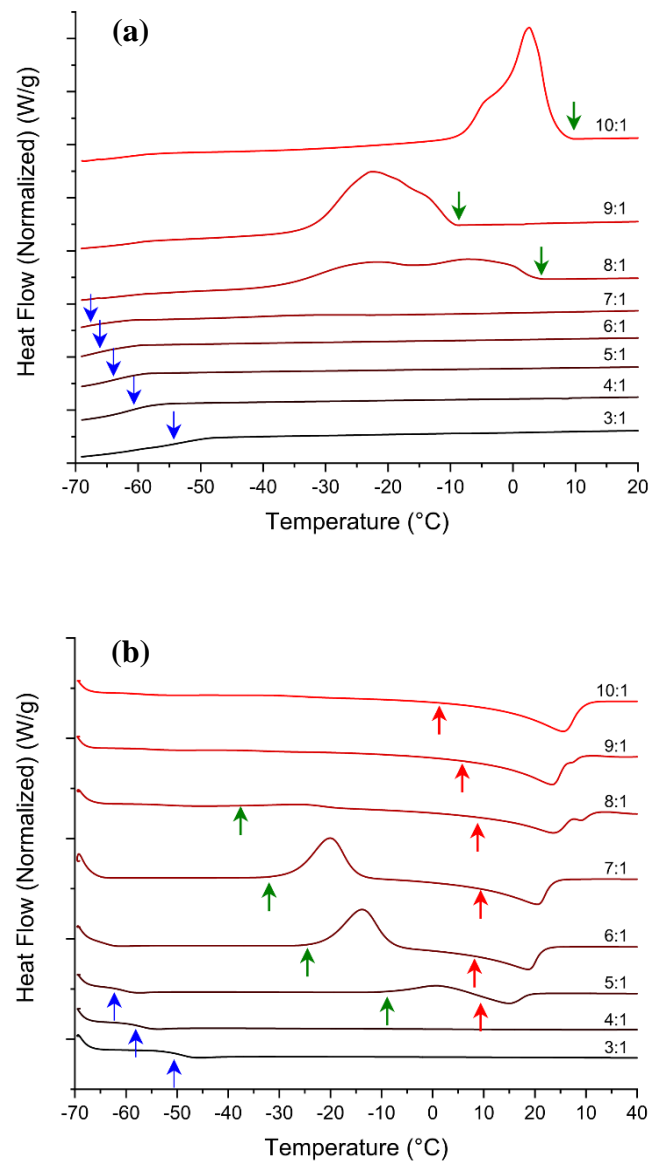

**Figure S10.** Thermal behaviour of metacetamol as a VODES with phenol. Thermograms of each VODES ratio for (a) cooling and (b) warming, truncated to remove featureless data. Coloured arrows of blue green and red indicate the positions of a $T_g$, a crystallisation event and a melt event, respectively.

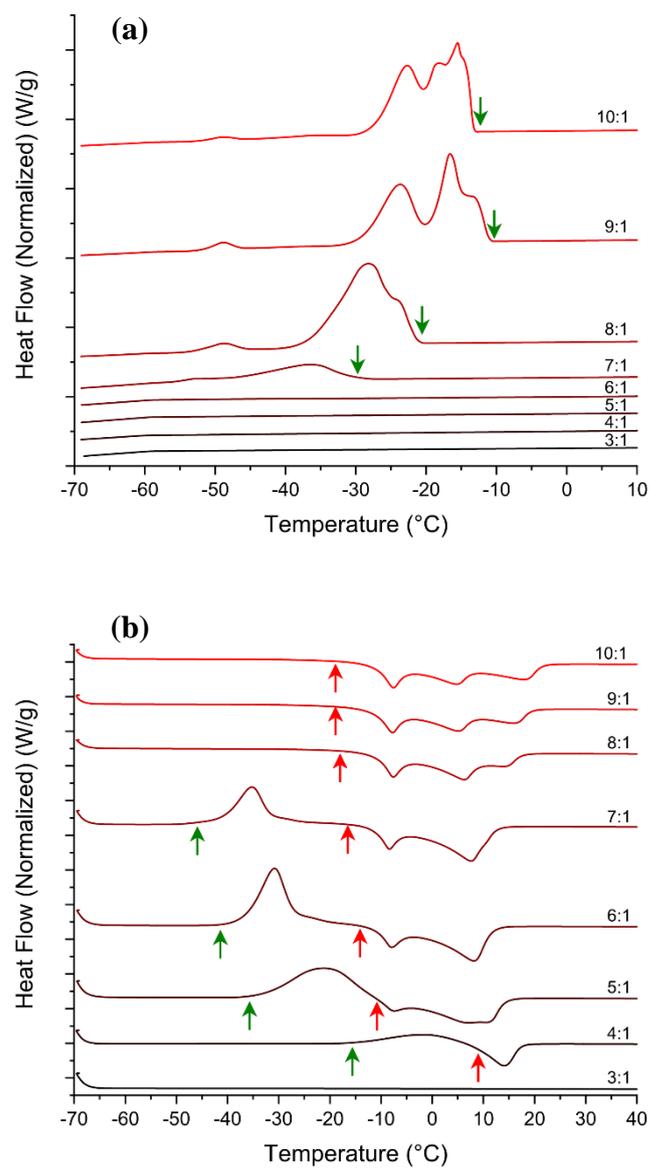

**Figure S11.** Thermal behaviour of benzamide as a VODES with phenol. Thermograms of each VODES ratio for (a) cooling and (b) warming, truncated to remove featureless data. Coloured arrows of blue green and red indicate the positions of a $T_g$, a crystallisation event and a melt event, respectively.

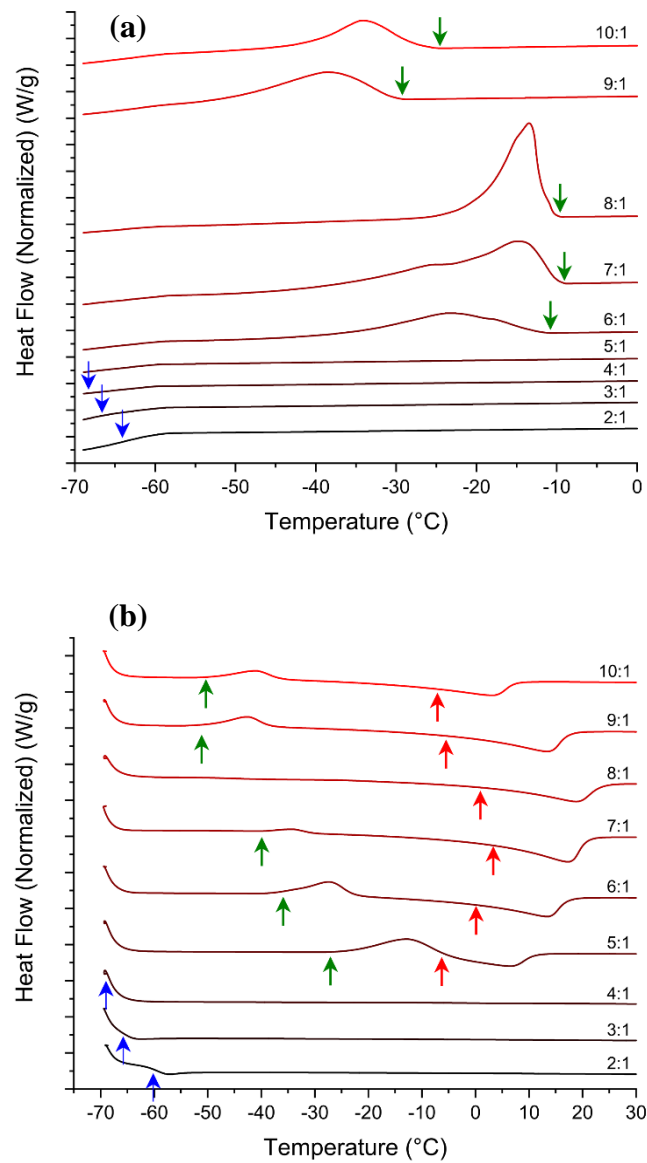

**Figure S12.** Thermal behaviour of 2MB as a VODES with phenol. Thermograms of each VODES ratio for (a) cooling and (b) warming, truncated to remove featureless data. Coloured arrows of blue green and red indicate the positions of a $T_g$, a crystallisation event and a melt event, respectively.

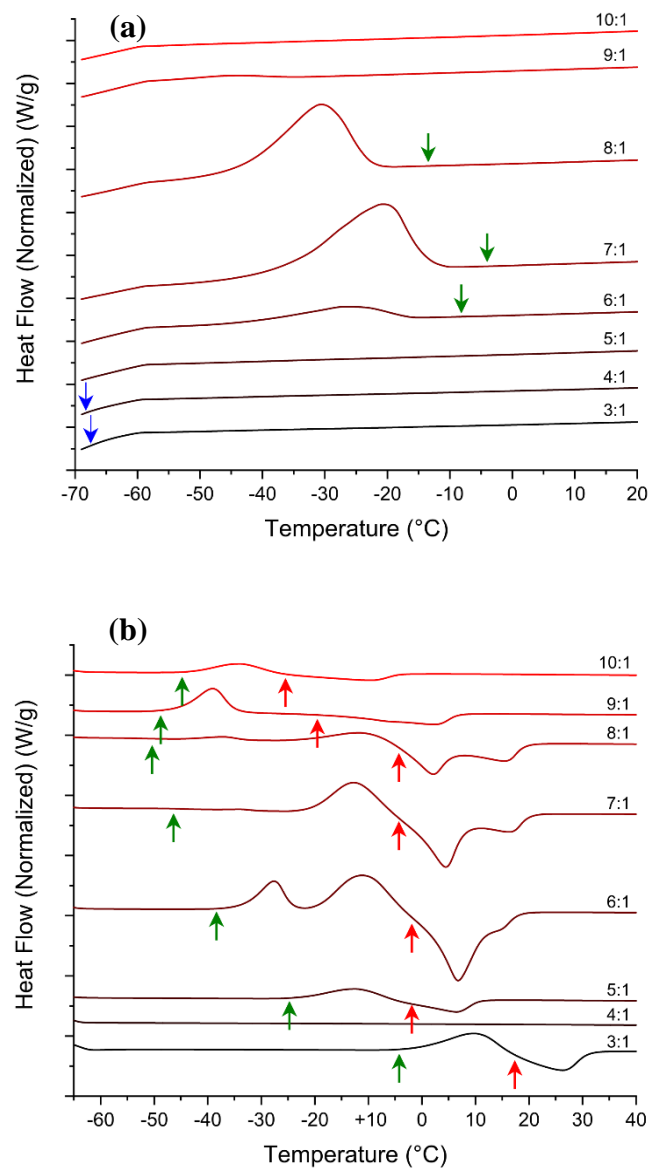

**Figure S13.** Thermal behaviour of 2EB as a VODES with phenol. Thermograms of each VODES ratio for (a) cooling and (b) warming, truncated to remove featureless data. Coloured arrows of blue green and red indicate the positions of a $T_g$, a crystallisation event and a melt event, respectively.

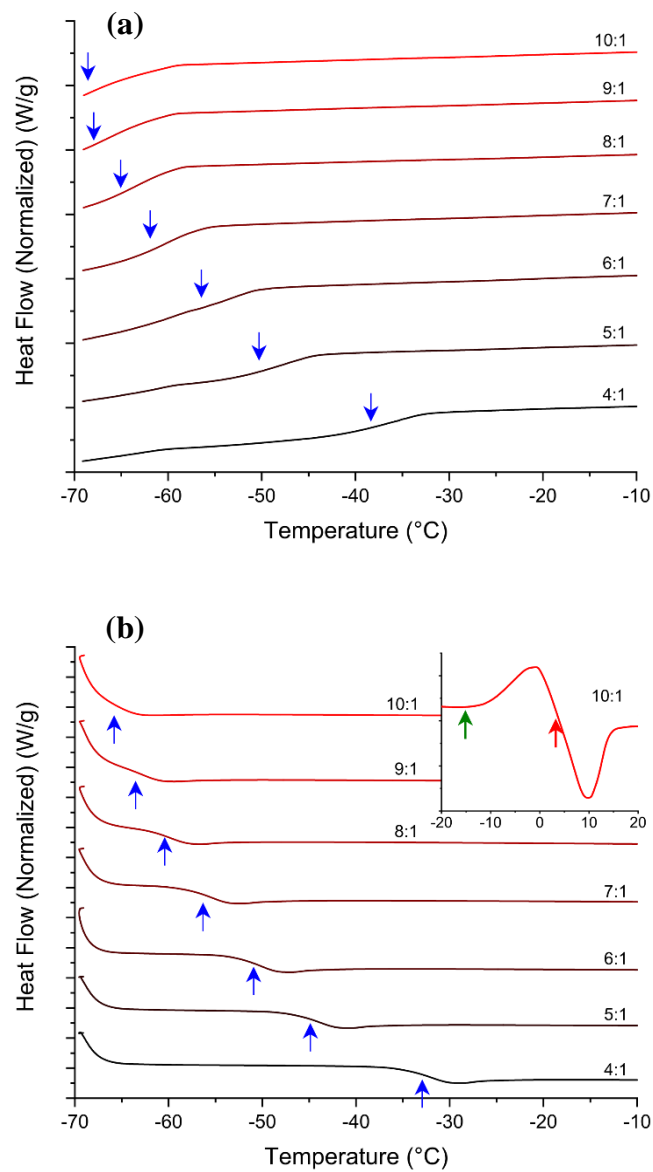

**Figure S14.** Thermal behaviour of carbamazepine as a VODES with phenol. Thermograms of each VODES ratio for (a) cooling and (b) warming, truncated to remove featureless data. Coloured arrows of blue green and red indicate the positions of a $T_g$, a crystallisation event and a melt event, respectively.

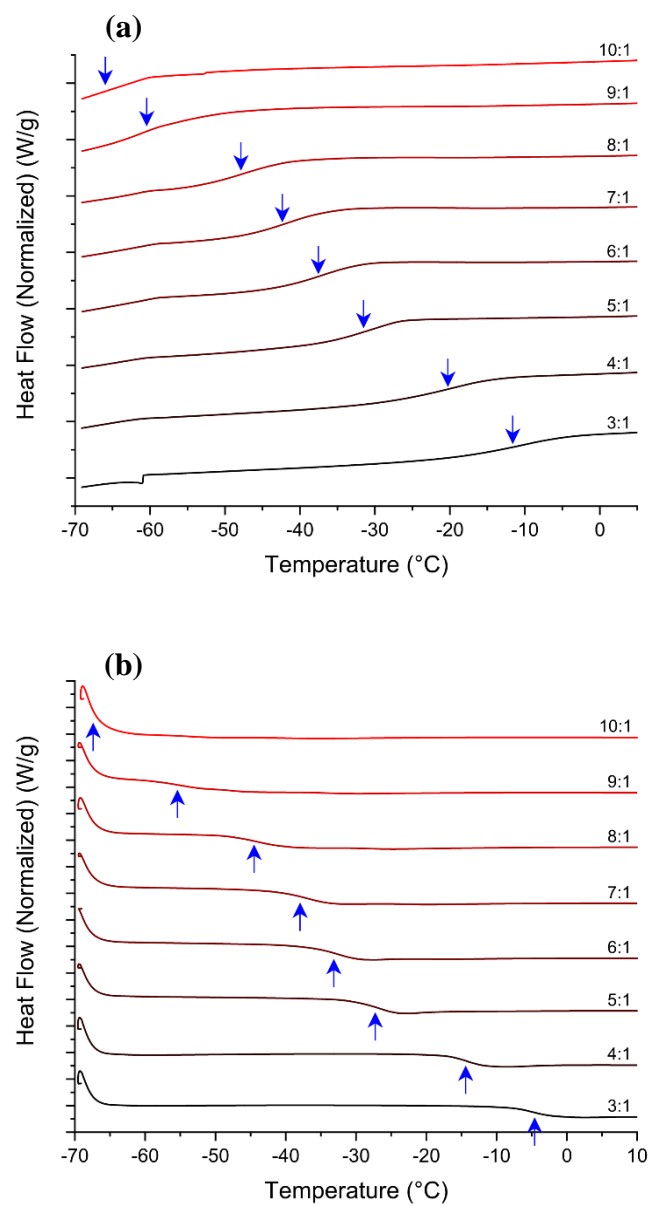

**Figure S15.** Thermal behaviour of verapamil hydrochloride as a VODES with phenol. Thermograms of each VODES ratio for (a) cooling and (b) warming, truncated to remove featureless data. Coloured arrows of blue green and red indicate the positions of a $T_g$, a crystallisation event and a melt event, respectively.

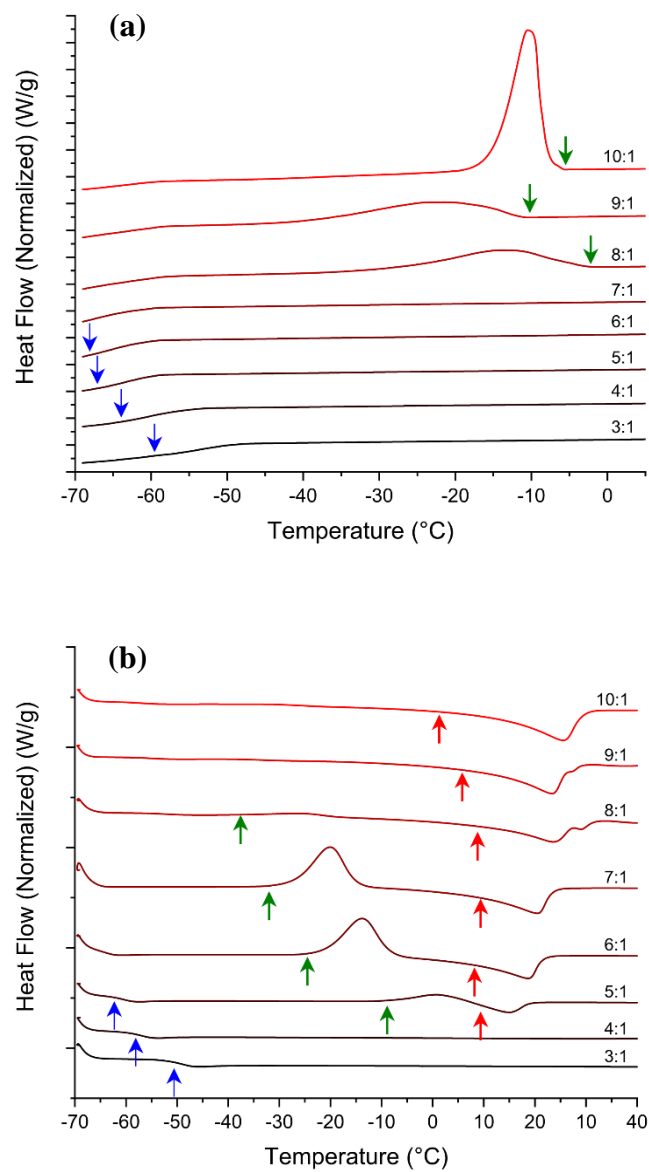

**Figure S16.** Thermal behaviour of metaxalone as a VODES with phenol. Thermograms of each VODES ratio for (a) cooling and (b) warming, truncated to remove featureless data. Coloured arrows of blue green and red indicate the positions of a $T_g$, a crystallisation event and a melt event, respectively.

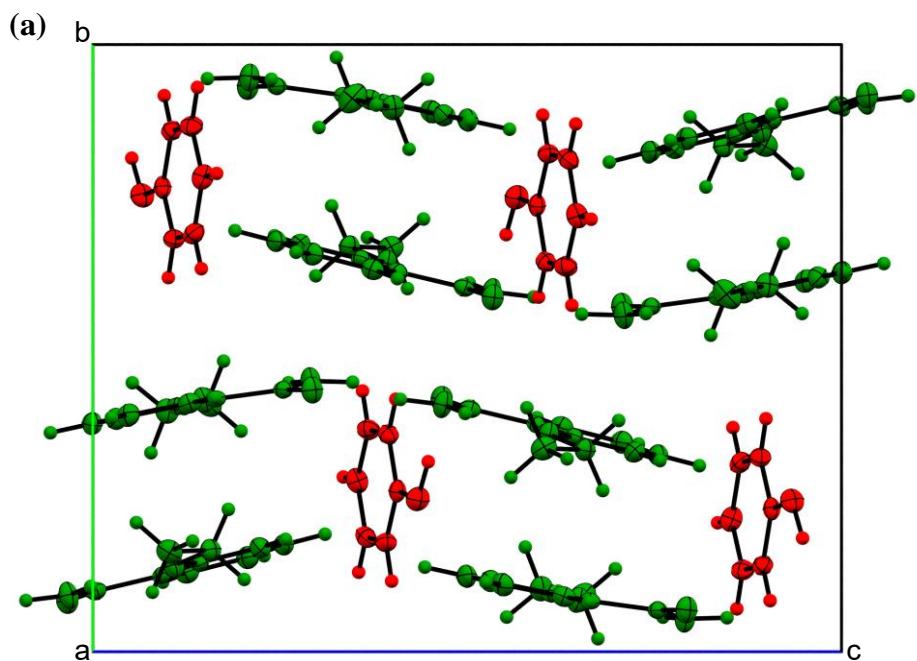

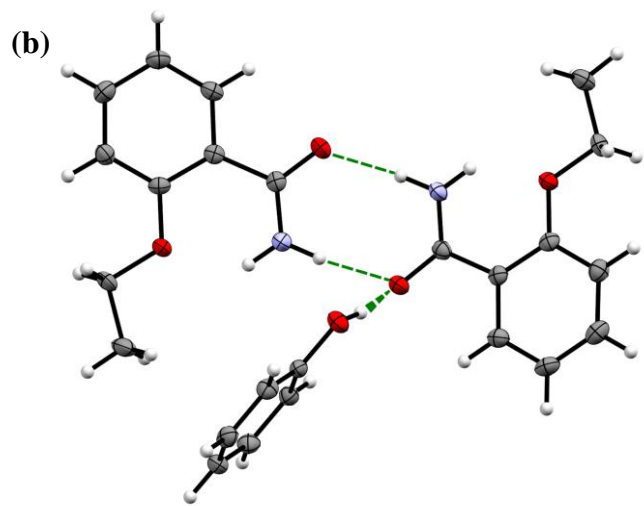

**Figure S17.** Crystal structure of a co-crystal of 2-ethoxybenzamide and phenol at a 1:2 ratio. (CCDC deposit number 1879336).

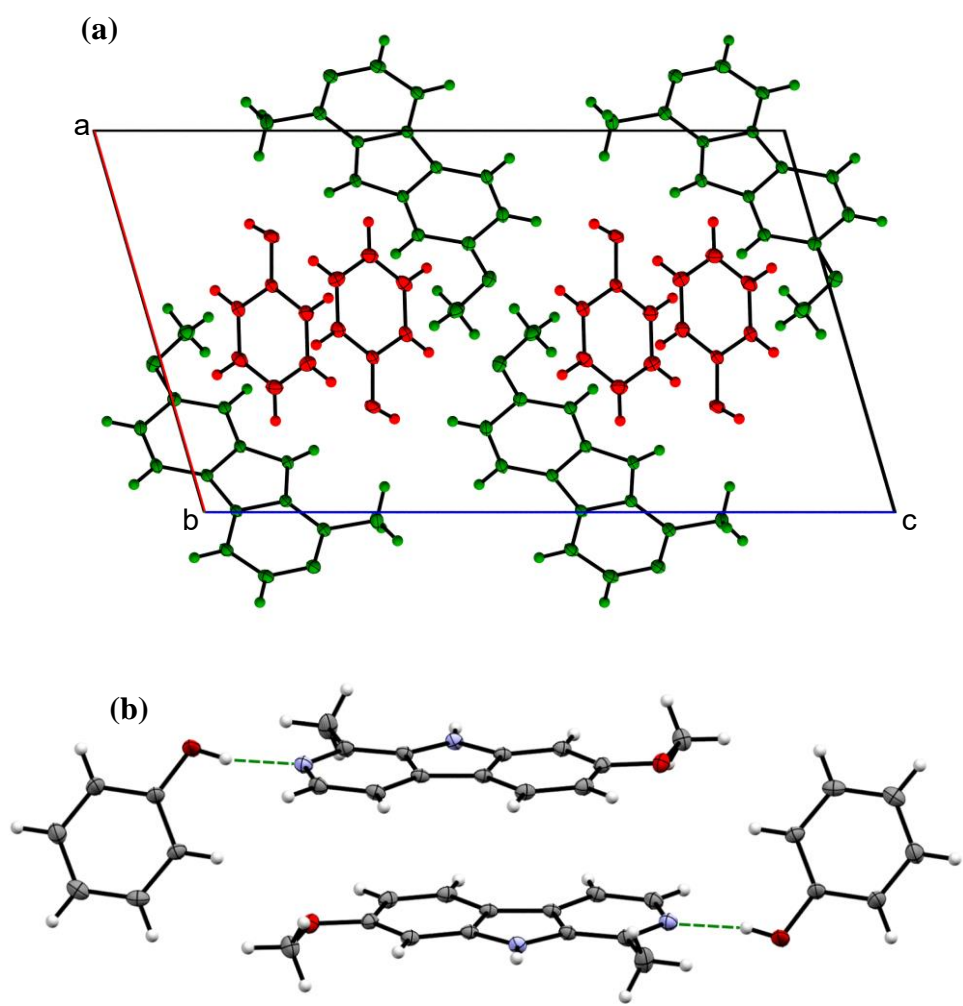

**Figure S18.** Crystal structure of a co-crystal of harmine and phenol at a 1:1 ratio. (CCDC deposit number 1879689).